\documentclass[12pt]{article}
\def\1{\'{\i}}
\def\3{\ss}

\usepackage{color,colortbl}

\definecolor{myaqua}{rgb}{0.0,0.5,0.55}
\definecolor{lightaqua}{rgb}{0.75,0.95,0.95}

\usepackage[colorlinks = true, 
  linkcolor = myaqua, 
  urlcolor  = blue, 
  citecolor = myaqua]{hyperref}

\usepackage{listings}

\newenvironment{allintypewriter}{\ttfamily}{\par}

\title{MHD Generation Code}
\author{Francisco Frutos Alfaro\thanks{School of Physics, 
University of Costa Rica, email:frutos@fisica.ucr.ac.cr} 
\\ Rodrigo Carboni M\'endez\thanks{School of Physics, 
University of Costa Rica, email:rcarboni@fisica.ucr.ac.cr}}
\date{\today}

\begin{document}
\maketitle

\abstract{
A program to generate codes in Fortran and C of the full Magnetohydrodynamic 
equations is shown. The program used the free computer algebra system software 
REDUCE. This software has a package called EXCALC, which is an exterior 
calculus program. The advantage of this program is that it can be modified 
to include another complex metric or spacetime. The output of this program 
is modified by means of a LINUX script which creates a new REDUCE program to 
manipulate the MHD equations to obtain a code that can be used as a seed for a 
MHD code for numerical applications. As an example, we present part of output 
of our programs for Cartesian coordinates and how to do the discretization.}

\section{Introduction \label{00}}

\noindent
Nowadays, there are programs that generate codes for a given mesh arrangement, 
but there is no a program to generate the differential equation in a given 
space-time. Now, we show how it is possible by means of a EXCALC package 
\cite{Schruefer} of REDUCE \cite{Hearn} and a LINUX script. REDUCE is 
a free computer algebra system (CAS) software intended to algebraic 
manipulation. In that sense, it is similar to Mathematica or Maple. 
The EXCALC package solves problems using the Cartan formalism or differential 
forms \cite{Arfken,Hassani}. 

\noindent
Due to the advent of new technology, there is an interest in plasma simulations 
and visualizations e.g. \cite{Carboni,Germaschewski,Pen}. 
There are a lot of information on these subjects in the Internet 
\cite{Birdsall,Jardin,Hockney,Hsu,Tajima}. Moreover, the mesh or grid 
generation is important in designing complex engineering structures or complex 
nonsymmetric systems \cite{Persson}.      

\noindent
We wrote a REDUCE program (\texttt{MHD.red}) and a long UNIX script 
(\texttt{dis\-cre\-tized-mhd}) mainly for Magnetohydrodynamics (MHD), but 
the program and script we present here are easy to modify for another purposes. 
Our program could be used together with a mesh generation code to solve complex 
problems.

\noindent
The paper is organized as follows. In section \ref{01}, a brief description of 
MHD is presented. A short summary of EXCALC is given in section \ref{02}. 
The explanation of the program is described in section \ref{03}. Moreover, 
we present the results of the discretization to Cartesian coordinates as an 
example. The conclusions and future work is discussed in section \ref{06}.

\section{MHD \label{01}}

\noindent
MHD is the study of the dynamics of electrically conducting fluids. 
Examples of such fluids include plasmas, liquid metals, and salt water or 
electrolytes. Plasmas can be regarded as fluids since the mean free paths for 
collisions between the electrons and ions are macroscopically long. Thus, 
collective interactions between large numbers of plasma particles can 
isotropize the particles velocity distributions in some local mean reference 
frame, thereby making it sensible to describe the plasma macroscopically by 
a mean density, velocity, and pressure. These mean quantities can then be shown 
to obey the same conservation laws of mass, momentum and energy for fluids.

\noindent
The fundamental concept behind MHD is that magnetic fields can induce currents 
in a moving conductive fluid, which in turn creates forces on the fluid and 
also changes the magnetic field itself. The set of equations which describe 
MHD are a combination of the Navier-Stokes equations of fluid dynamics and 
Maxwell's equations of electromagnetism. 

\noindent
An important physical parameter that defines the way MHD treats the systems is 
resistivity.  When  the resistivity is very low plasmas can be treated as 
perfect conductors (ideal MHD in the infinite magnetic Reynolds number limit) 
and the the lines of force appear to be dragged along with the conductor.

\noindent
When the resistivity cannot be neglected the magnetic field can generally move 
through the fluid following a diffusion law with the resistivity of the plasma 
serving as a diffusion constant. According to this the solutions to the ideal 
MHD equations are only applicable for a limited time before diffusion becomes 
too important to ignore. The diffusion time across a solar active region is 
hundreds to thousands of years. The interested reader may consult the 
references \cite{Cap,Dendy,Jackson,Kulikovskiy}.

\noindent
There are many applications of MHD but in this article we are interested in 
those related to astrophysics and cosmology. Plasma is the constituent of 
many astrophysical objects of the Universe: stars, nebulae, the interplanetary, 
the interstellar and the intergalactic media.

\noindent
Many astrophysical problems require the treatment of magnetized fluids in 
dynamical, strongly curved spacetimes. Such problems include the origin of 
gamma-ray bursts, magnetic braking of differential rotation in incipient 
neutron stars arising from stellar core collapse or binary neutron star merger, 
the formation of jets and magnetized disks around newborn black holes and many 
others \cite{Tajima2}.

\bigskip
\noindent
{\bf MHD Equations}:
\bigskip

\noindent
{\bf Continuity Equation}

\noindent
The conservation of mass in a fluid tell us that

\begin{equation}
\label{continuity}
\frac{\partial \rho}{\partial t} + \nabla \cdot {\bf J}_s = 0 .
\end{equation}

\noindent
where $\rho $ is the density, $ {\bf v} $ is the fluid velocity and 
$ {\bf J}_s = \rho {\bf v} $.

\bigskip
\noindent
{\bf Maxwell Equations}

\noindent
The electric Gau{\3} law is 

\begin{equation}
\label{gauss1}
\epsilon_0 \nabla \cdot {\bf E} = \rho_e ,
\end{equation}

\noindent
where $ {\bf E} $ is the electric field, $ \rho_e  $ is the electric charge 
density and $ \epsilon_0 $ is the vacuum permittivity.

\noindent
The magnetic Gau{\3} law is 

\begin{equation}
\label{gauss2}
\nabla \cdot {\bf B} = 0 .
\end{equation}

\noindent
where $ {\bf B} $ is the magnetic field.

\noindent
The Amper\'e law is 

\begin{equation}
\label{ampere}
\nabla \times {\bf B} = \mu_0 {\bf J} 
+ \mu_0 \epsilon_0 \frac{\partial {\bf E}}{\partial t} .
\end{equation}

\noindent
where $ {\bf J} $ is the electric current density and $ \mu_0 $ is 
the vacuum permeability. 

\noindent
The Faraday law is 

\begin{equation}
\label{faraday}
\nabla \times {\bf E} = - \frac{\partial {\bf B}}{\partial t} .
\end{equation}

\bigskip
\noindent
{\bf Equation of Motion}

\noindent
The equation of motion of a fluid or the Navier-Stokes equation with 
electromagnetic fields is

\begin{equation}
\label{motion}
\frac{\partial {\bf J}_s}{\partial t} = {\bf F}_L 
+ \nabla \cdot [{\bf P} - \rho {\bf v}{\bf v}] .
\end{equation}

\noindent
where $ {\bf P} $ is a the stress tensor with components

\begin{equation}
\label{pij}
P_{i j} = - P \delta_{i j} + {\cal T}_{i j} 
\end{equation}

\noindent
$ P $ is the pressure and $ {\cal T}_{i j} $ represents the viscosity stress 
tensor components, which are given by

\begin{equation}
\label{tij}
{\cal T}_{i j} = \mu \left( \frac{\partial v_i}{\partial x_j} 
+ \frac{\partial v_j}{\partial x_i} 
- \frac{2}{3} \nabla \cdot {\bf v} \delta_{i j} \right) 
+ \nu \nabla \cdot {\bf v} \delta_{i j} , 
\end{equation}

\noindent
with $ \mu $ and $ \nu $ are the first and second viscosity coeficients.

\noindent
$ {\bf F}_L $ is the volume Lorentz force given by  

\begin{eqnarray}
\label{lorentz}
{\bf F}_L & = & \rho_e {\bf E} + {\bf J} \times {\bf B} \\
& = & \nabla \cdot {\bf T} 
- \epsilon_0 \mu_0 \frac{\partial {\bf S}}{\partial t} \nonumber 
\end{eqnarray}

\noindent
where $ {\bf S} $ is the Poynting vector

\begin{equation}
\label{poynting}
{\bf S} = \frac{1}{\mu_0}{\bf E} \times {\bf B} .
\end{equation}

\noindent
and $ {\bf T} $ is the electromagnetic tensor with following components

\begin{equation}
\label{emtensor}
T_{ij} = {\epsilon_0} E_i E_j + \frac{1}{\mu_0} B_i B_j 
- \left({\epsilon_0} \frac{E^2}{2} + \frac{B^2}{2 \mu_0} \right) \delta_{ij}
\end{equation}

\bigskip
\noindent
{\bf The Ohm Law}

\noindent
The Ohm law is

\begin{equation}
\label{ohm}
{\bf E} = \eta {\bf J} - {\bf v} \times {\bf B} ,
\end{equation}

\noindent
$ \eta $ is the resistivity.

\bigskip
\noindent
{\bf Equation of State}

\noindent
The equation of state for an ideal gas is

\begin{equation}
\label{ideal}
{P} = \rho R T .
\end{equation}

\noindent
where $ T $ is the temperature and $ R $ is the gas constant.

\bigskip
\noindent
{\bf Equation of Energy Conservation}

\noindent
The equation of energy conservation is

\begin{equation}
\label{energycon}
\frac{\partial}{\partial t} \left[\rho \left(
\epsilon + \frac{v^2}{2} \right) \right] = - \left[\nabla \cdot {\bf Q} 
- {\bf J} \cdot {\bf E} \right] .  
\end{equation}

\noindent
where 

\begin{equation}
\label{vecq}
{\bf Q} = \left(\epsilon + \frac{v^2}{2} \right) {\bf J}_s + {\bf q} 
- {\bf P} \cdot {\bf v} 
\end{equation}

\noindent
and $ \epsilon $ is the internal energy 

\begin{equation}
\label{intenergy}
\epsilon = C_V T = \frac{P}{(\gamma-1) \rho} ,
\end{equation}

\noindent
here $ \gamma $ is the adiabatic exponent

\begin{equation}
\label{adiabaticexp}
\gamma = \frac{C_P}{C_V} = \frac{R+C_V}{C_V} , 
\end{equation}

\noindent
and $ {\bf q} $ is the heat flux vector

\begin{equation}
\label{fourier}
{\bf q} = - k \nabla T ,
\end{equation}

\noindent
with $ k $ is the thermal conductivity.

\section{EXCALC \label{02}}

\noindent
EXCALC is a package under REDUCE intended to solve problems with Cartan 
exterior calculus. The advantage of using it, is that given equations can be 
written independent on the coordinate system. The outcomes of the calculations 
from these equations are expressed in the chosen coordinate system. 
Here, it is a list of the most used commands:

\smallskip
\begin{tabular}{ll}
{\^{}}           & Exterior multiplication \\
{\texttt{@}}     & Partial differentiation \\ 
{\#}             & Hodge $ \star $ operator \\
{\_$|$}          & Inner product \\
{$|$\_}          & Lie derivative \\
\texttt{COFRAME} & Declaration of a coframe \\ 
\texttt{d}       & Exterior differentiation \\
\texttt{FDOMAIN} & Declaration of implicit dependencies \\
\texttt{FRAME}   & Declares the frame dual to the coframe \\
\texttt{METRIC}  & Clause of \texttt{COFRAME} to specify a metric \\ 
\texttt{PFORM}   & Declaration of exterior forms \\
\end{tabular}

\noindent
The principal operations with 1-forms or vectors are listed below

\smallskip
\begin{tabular}{ll}
Cross product & \#(\texttt{v}{\^{}}\texttt{B}) 
(\texttt{B} and \texttt{v} are vectors or 1-forms) \\
Nabla      & \texttt{d P} (\texttt{P} is a function or scalar) \\
Divergence & {\#}\texttt{d}{\#} \texttt{F} (\texttt{F} is a vector or 1-form) \\
Curl       & {\#}\texttt{d F} (\texttt{F} is a vector or 1-form)
\end{tabular}

\noindent
For more information about this REDUCE package, the interested user may 
consult the references.

\section{The Program \label{03}}

\noindent
The REDUCE program \texttt{MHD.red} to transform the MHD equations into 
Cartesian coordinates is included as an appendix. The script 
\texttt{discretized-mhd} and the generating code are not include in this paper, 
but it can be emailed send by the authors if requested, and in a near future 
will be available at our {\it Space Research Center} 
Webpage\href{http://cinespa.ucr.ac.cr/software/}{
\color{blue}{http://cinespa.ucr.ac.cr/software/}}.

\subsection{Example of the Script for Cartesian Coordinates \label{04}}

\noindent
Now, we will present the modified output of the programs for a given 
MHD equations using an UNIX script. Let us consider the induction equation:

$$ \frac{\partial {\bf B}}{\partial t} + \nabla\times {\bf E} , $$

\noindent
where $ {\bf E} = \eta {\bf J}_m - {\bf v} \times {\bf B} $ and 
$ \mu_0 {\bf J}_m = \nabla \times {\bf B} $, without taking into account the 
time variation of the electric field.

\noindent
For example, the outcome from our program for the $ y $ component of 
the induction equation (Car\-te\-sian coordinates) is 

\smallskip
\begin{allintypewriter}
{$\!\!\!\!\!\!\!\!\!$}induct( - x2) := @(b( - x2),t) \\ 
- b( - x1)*@(v( - x2),x1) + b( - x2)*@(v( - x1),x1) \\
+ b( - x2)*@(v( - x3),x3) - b( - x3)*@(v( - x2),x3) \\
- @(b( - x1),x1)*v( - x2) + @(b( - x2),x1)*v( - x1) \\
+ @(b( - x2),x3)*v( - x3) - @(b( - x3),x3)*v( - x2) \\
+ (@(b( - x1),x1,x2)*eta - @(b( - x2),x1,x1)*eta \\
+ @(b( - x1),x2)*@(eta,x1) - @(b( - x2),x1)*@(eta,x1) \\ 
- @(b( - x2),x3)*@(eta,x3) + @(b( - x3),x2)*@(eta,x3) \\
- @(b( - x2),x3,x3)*eta + @(b( - x3),x2,x3)*eta)/mu\_0
\end{allintypewriter}

\smallskip
\noindent
As this output is not easy to handle, we translate such equation to a 
FORTRAN or C code, then we wrote a script that modify this outcome into 
equations with a prescribed discretization method. To this aim, 
we employ the UNIX command \texttt{sed} \cite{Unix}. This command change the 
output using the prescribed discretization method. For instance, the following 
commands 

\smallskip
\noindent
\begin{allintypewriter}
{$\!\!\!$}sed -e 's:@(b( - x1),t):(bx(n+1,i,j,k)-bx(n,i,j,k))/dt:g' \\ 
mhdeqs r mhdeqsnew \\
sed -e 's:@(v( - x2),x1):(vy(n,i+1,j,k)-vy(n,i-1,j,k) \\ 
)/(2*dx):g' mhdeqs r mhdeqsnew \\
\end{allintypewriter}

\noindent
will produce the translation

$$ \frac{\partial b_x}{\partial t} \longrightarrow 
\frac{b_x(n+1,\, i, \, j, \, k) - b_x(n,\, i, \, j, \, k)}{\Delta t} $$

$$ \frac{\partial v_y}{\partial x} \longrightarrow 
\frac{v_x(n,\, i + 1, \, j, \, k) - v_x(n,\, i - 1, \, j, \, k)}{2 \Delta x} 
. $$

\noindent
The output of our REDUCE program is \texttt{mhdeqs} (see Appendix) and 
the \texttt{mhdeqsnew}, which should be created before one runs the script, 
will contain all modifications of the entire system of equations.

\subsection{Example of the Code Generation \label{05}}

\noindent
Here, we show a part of the output after the script is run. In this case, 
the $ y $ component of the induction equation is converted into 
 
\smallskip
\noindent
\begin{allintypewriter}
{$\!\!\!$}induct\_y := (by(n+1,i,j,k)-by(n,i,j,k))/dt \\
+ ((bx(n,i+1,j+1,k)-bx(n,i+1,j-1,k) \\
-bx(n,i-1,j+1,k)+bx(n,i-1,j-1,k))/(4*dx*dy)*eta(n,i,j,k) \\ 
+ (bx(n,i,j+1,k)-bx(n,i,j-1,k))/(2*dy)*(eta(n,i+1,j,k) \\ 
-eta(n,i-1,j,k))/(2*dx) \\
- (by(n,i+2,j,k)-2*by(n,i,j,k)+by(n,i-2,j,k))/(4*dx*dx)* \\ eta(n,i,j,k) \\
- (by(n,i+1,j,k)-by(n,i-1,j,k))/(2*dx)*(eta(n,i+1,j,k) \\ 
-eta(n,i-1,j,k))/(2*dx) \\
- (by(n,i,j,k+2)-2*by(n,i,j,k)+by(n,i,j,k-2))/(4*dz*dz)* \\ 
eta(n,i,j,k) \\
- (by(n,i,j,k+1)-by(n,i,j,k-1))/(2*dz)*(eta(n,i,j,k+1) \\ 
-eta(n,i,j,k-1))/(2*dz) \\
+ (bz(n,i,j+1,k+1)-bz(n,i,j+1,k-1) \\
-bz(n,i,j-1,k+1)+bz(n,i,j-1,k+1))/(4*dy*dz)*eta(n,i,j,k) \\
+ (bz(n,i,j+1,k)-bz(n,i,j-1,k))/(2*dy)*(eta(n,i,j,k+1) \\
-eta(n,i,j,k-1))/(2*dz))/mu\_0
\end{allintypewriter}

\noindent
Thus, it is possible to use this output as a FORTRAN or C code.

\section{Conclusion \label{06}}

\noindent
In this paper, we show how to produce quickly a computer code useful 
to implement simulations or visualizations. The automatic generation of the
computer code is possible by means of REDUCE programs and UNIX scripts that 
modify and manipulate the output of the programs and generate computer codes 
with a given discretization method. The programs shown here, are flexible and 
are easy to modify for other purposes.

\noindent
Due to the rapid advance of technology, there is an interest in computer
simulations and visualizations, in particular in plasma physics. 
Our program together with recent developed grid generation codes helps 
solve complex plasma phenomena.

\section{Appendix}


\begin{allintypewriter}
\% REDUCE Program MHD.red \\
\% It is based on an example of E. Schruefer \\
out mhdeqs\$ \\
load excalc\$ \\
off nat\$ \\
\% Problem: \\
\% \\
\% Calculate in spherical coordinates system the MHD \\ 
\% equations. \\
\% coframe e r = d r, e theta = r*d theta, \\
\% e phi = r*sin(theta)*d phi\$ \\ 
\% frame x\$ \\
\% fdomain v = v(t, r, theta, phi), \\ 
\% p = p(t, r, theta, phi)\$ \\
\% \\
\% Calculate in cartesian coordinates system the MHD \\ 
\% equations. \\
coframe e x1 = d x1, e x2 = d x2, e x3 = d x3\$ \\
frame x\$ \\
fdomain v=v(t,x1,x2,x3), p=p(t,x1,x2,x3)\$ \\
fdomain rho=rho(t,x1,x2,x3), je=je(t,x1,x2,x3)\$ \\
fdomain pstar=pstar(t,x1,x2,x3), b=b(t,x1,x2,x3)\$ \\
fdomain eta=eta(t,x1,x2,x3), energy=energy(t,x1,x2,x3)\$ \\
fdomain jm=jm(t,x1,x2,x3), jmp=jmp(t,x1,x2,x3)\$ \\
fdomain ee=ee(t,x1,x2,x3), a=a(t,x1,x2,x3)\$ \\
pform v(k)=0, je(k)=0, p=0, rho=0\$ \\
v := v(-k) * e(k); \\
\% je := rho * v; \\
je := je(-k) * e(k); \\
\%factor e, @\$ \\
factor e\$ \\
on rat\$ \\
\% \\
\% First we calculate the continuity equation. \\
\% drho/dt + nabla.je = 0 \\
pform conteq=0\$ \\
conteq := @(rho,t) + \#d\# je; \\
\% \\
\% Next we calculate the equation of motion. \\
\% d je/dt + nabla pstar + (nabla.je) v + je.nabla v  \\
\%         + (nabla.b) b + b.nabla b = 0 \\
\% nabla.b = 0 \\
pform moveq=1, moveq(-k)=0, pstar=0, b(-k)=0\$ \\
b := b(-k) * e(k); \\
\% pstar := p+(b(k) * b(-k))/(2*mu\_0); \\
\% moveq := @(je,t) + d pstar + (\#d\# je)*v \\
\% + rho*((v(k) * x(-k)) |\_ v - (1/2)*d (v(k) * v(-k))) \\ 
\% + (\#d\# b)*b + ((b(k) * x(-k)) |\_ b \\
\% - (1/2)*d (b(k) * b(-k))); \\
moveq := @(je,t) + d pstar + (\#d\# je)*v \\
+ rho*((v(k) * x(-k)) |\_ v - (1/2)*d (v(k) * v(-k))) \\ 
+ ((b(k) * x(-k)) |\_ b - (1/2)*d (b(k) * b(-k))); \\
moveq(-k) := x(-k) \_| moveq; \\
\% \\
\% Next we calculate Gauss equation for the magnetic \\ 
\% field. \\
\% nabla.b = 0 \\
pform gauss=0; \\
gauss := \#d\# b; \\
\% \\
\% Next we calculate the induction equation. \\
pform induct=1, induct(-k)=0, ee=1, ee(-k)=0, \\ 
jm=1, jm(-k)=0, eta=0\$ \\
pform jmp=1, jmp(-k)=0, jmpp=1, jmpp(-k)=0, \\
a=1, a(-k)=0, bb=1, bb(-k)=0\$ \\
pform induct2=1, induct2(-k)=0, ee2, ee2(-k)=0\$ \\
jm := (\#d b)/mu\_0; \\
jm(-k) := x(-k) \_| jm; \\
\% b := \#d a; \\
a := a(-k) * e(k); \\
jmp:= jmp(-k) * e(k); \\
bb := \#d a; \\
jmpp := (\#d bb)/mu\_0 - jmp; \\
jmpp(-k) := x(-k) \_| jmpp; \\
ee := eta*jm - \#(v{\^{}}b); \\
ee(-k) := x(-k) \_| ee; \\
ee2 := eta*jm - \#(v{\^{}}bb); \\
ee2(-k) := x(-k) \_| ee2; \\
induct := @(b,t) + \#d ee; \\
induct(-k) := x(-k) \_| induct; \\
induct2 := @(a,t) + eta*(\#d bb)/mu\_0 - \#(v{\^{}}bb); \\
induct2(-k) := x(-k) \_| induct2; \\
\% \\
\% Finally we calculate the energy conservation. \\
\% energy := rho * v**2/2 + p/(gamma-1) + b**2/(2*mu\_0); \\
pform energy=0, enconser=0\$ \\
\% energy := rho * (v(k) * v(-k))/2 + p/(gamma-1) \\
\% + (b(k) * b(-k))/(2*mu\_0); \\
enconser := @(energy,t) + \#d\# ((energy+pstar)*v \\ 
- ((v(k) * x(-k)) \_| b)*b); \\
\% \\
clear v, x, b, p, pstar, eta, conteq, moveq, gauss, \\ 
ee, energy, enconser\$ \\
remfac e, @\$ \\
remfdomain p, v, b, energy, eta, pstar, rho, je\$ \\
shut mhdeqs\$ \\
bye; 
\end{allintypewriter}


\end{document}